\documentclass[usegraphicx]{mn2e}
\begin{document}

\title[Gas disc collapse]{The collapse of gas discs in non-axisymmetric 
  galaxy cores}

\author[R.H. Sanders] {R.H.~Sanders\\Kapteyn Astronomical Institute,
P.O.~Box 800,  9700 AV Groningen, The Netherlands}

 \date{received: ; accepted: }

\maketitle

\begin{abstract}
Below a threshold energy, gas in the constant density core of a 
triaxial galaxy can find no simple non-intersecting periodic orbit
to act as an attractor for its trajectory (El-Zant et al. 2003).  
If a disc of gas arriving from
further out in the galaxy dissipates sufficient energy to
fall below this threshold, it will thereafter collapse into the very centre.
Such a mechanism may be relevant to the early growth of super-massive
black holes at the Eddington limit and the appearance of the
quasar phenomenon at high redshift.
This process is self-limiting in the sense that,
when the black hole mass has grown to a significant fraction of the
core mass, simple angular momentum conserving orbits are restored
and accretion reverts to the slow viscous mode.
The mechanism depends upon the pre-existence of
constant density cores in triaxial spheroidal galaxies.
\end{abstract}


\section{Introduction}

It is now evident that the nuclei of most, if not all, spheroidal galaxies 
contain super-massive black holes (SBH). The fact that the phenomenon
of extreme activity in galactic nuclei-- the quasar phenomenon-- 
peaked at early cosmic epochs and declined subsequently (e.g. Wall
et al. 2005)
suggests that the construction of
massive black holes in galaxies was coeval with galaxy formation
itself-- that black hole formation is an integral part of the galaxy
formation process.  The tight correlation between black hole
mass and the galaxy stellar velocity dispersion-- the
$M_h-\sigma$ relation-- \cite{fm,nuk1}, 
which requires a close connection between the black hole and the larger
scale galaxy kinematics, would seem to support this idea of 
simultaneous growth.

An alternative (and older) viewpoint is that galaxies,
and the corresponding deep potential wells, were in place essentially
before the black holes began their rapid growth-- that preexisting
galactic centres defined the sites for SBH formation 
rather than vice versa (see Merritt, 2006,
for a recent summary and a more complete list of references).
If true, then the epoch of galaxy formation must be further
back in cosmic history ($z\ge 10$) than the appearance of the 
quasar phenomenon.
The recent discovery of high redshift
galaxies \cite{hred1,hred2} would be consistent this view.  
The $M_h-\sigma$ scaling relation would then appear to require that
some aspect of galaxy structure should both
promote and then limit the growth of the black hole.  This is the
possibility that I will explore here in the context of a specific
mechanism for supplying the matter necessary for black hole growth. 

The mass source necessary for the rapid growth of nuclear black holes--
the construction material-- and its transport to the
near vicinity of the event horizon
have always been problematic.   Of course, an important mechanism for the
growth to high mass could well be the merging of preexisting black holes due
to galaxy mergers.  However, the quasar phenomenon itself and
the contribution of active nuclei to the X-ray background 
\cite{elvis,fabiw} would seem to
require that a substantial fraction of the present mass content
in SBHs is due growth from lower mass seeds 
via accretion of gas at the Eddington limit, with a high
efficiency of producing electromagnetic radiation. 

Diffuse gas in spheroidal galaxies is an obvious building material
for super-massive black holes.  It may arrive in the central regions
of a galaxy from either internal or external sources
\cite{shlos}, and unlike stars, 
gas is dissipative; through energy loss, the gas can sink deeper into 
the potential well.  However, for the Eddington growth of black holes,
rapid loss of angular momentum is a more significant problem.  To arrive
at the core of a galaxy (at radii of 10-1000 pc) gas initially
in equilibrium further out (at roughly an
effective radius) must loose typically 90\% of its angular momentum.
This appears to be possible due to the effects 
of non-axisymmetric distortions
of the potential-- intrinsic bars, or bars excited during encounters and
mergers (e.g. Sellwood \& Moore 1999).

But an even more significant problem is transport from the galaxy
core to the Schwarzschild radius, where the gas must reduce its
angular momentum by a factor of $10^7-10^8$.  The viscous
inflow timescale in a classical
accretion disc with subsonic turbulence appears to be
much too long
to fuel an extended period of high-luminosity activity \cite{shlos}. 
Therefore, here I consider a simple 
mechanism which may bridge this final, but significant, gap. To work,
this mechanism requires two assumptions:  first, that
spheroidal systems, such as elliptical galaxies or bulges of spiral 
galaxies,
initially contain a constant density core at their centres; and second,  that 
these cores are mildly non-axisymmetric.  

The first assumption is contentious.  The prevailing
viewpoint at present is that power law cusps form naturally in 
dissipationless collapse and constant density cores may be created
later by the ``scouring'' action of binary SBHs (i.e., gravitational
scattering of
stars by two in-spiralling black holes \cite{mm}). This idea
is supported by the fact that higher luminosity spheroidal
systems, those in which merging of equal partners
has probably played a more dominant role,
often appear to contain constant density cores; whereas, lower mass systems,
such as the Milky Way bulge, are power law ($r^{-1.5}$) into small radii
\cite{fabereal,merrev}.  On the other hand, evolution could go
the other way:  Cores may originally be present in galactic nuclei and
then altered by the growth of the SBH and the resulting
adiabatic or diffusive formation of a cusp in the stellar density
distribution \cite{peeb,bw,yng}.  
The appearance of the original core may also be altered as a 
consequence of subsequent star
formation via processes such as the one I will consider here. 

The second assumption is less controversial.  For some years now
it has been appreciated that slowly rotating but non-spherical
hot stellar systems are supported by an anisotropic velocity 
distribution.  This provides naturally a triaxial system
with a non-rotating figure \cite{schw}.  Here we require a gravitational
potential which is mildly non-axisymmetric in a principal plane.  This
asymmetry must extend into the core, and, indeed, various 
indicators of non-axisymmetric structures are
actually observed in galactic nuclei \cite{shaw}. 

Given these two assumptions, the mechanism is
simple: a gas disc is supported against gravity
by motion on near-circular orbits-- orbits which serve as
the parents of the tube families.  But within a constant density
non-axisymmetric core there are no tube orbits below a 
critical orbital energy; there are only
box orbits which, after sufficient time, pass arbitrarily close to the
centre (see e.g. Binney \& Tremaine 1987).  That is to say, within such
a core, orbital angular momentum-- or even a sense of rotation-- 
is not conserved; the two integrals
or motion are effectively the oscillation periods along the
principal axes, and, if these are non-commensurate, the 
orbit fills an elongated box after infinite time.  Then
the gas, being dissipative, accumulates at the centre.

This idea is not new.  It was originally suggested by 
Lake \& Norman (1983) in a
wide-ranging paper discussing the orbit structure in triaxial
systems and the relationship between that orbit structure and
gas flow.  At about this time it was appreciated that simple
non-self-intersecting periodic orbits act as attractors for
gas flow in non-axisymmetric systems, and much of the structure
in gas-rich galaxies can be understood in the context of this
fact \cite{stva}.  Lake and Norman realized that if there are no
simple periodic orbits over some range of energy-- that if there
is no integral of motion preserving a sense of rotation like 
angular momentum-- then the ultimate
attractor is the centre of the galaxy, and one might expect gas
inflow to be significantly enhanced.

The inability of a gas disc to be sustained in a constant density
non-axisymmetric core is the central aspect of a model by El-Zant et al. 
(2003) for 
the simultaneous formation of a SBH and axisymmetric spheroid in the
presence of a triaxial CDM halo.  A constant density
core is presumably created in
the cuspy CDM halo by the scouring action of baryonic clumps \cite{ezsh}.
The re-emergence of tube orbits in the increasingly axisymmetric
potential would then limit the growth of both the SBH and the
spheroid and, with additional assumptions, explain the observed 
$M_h-\sigma$ 
relation.  Here I take the standpoint that the baryonic component
of a spheroidal galaxy is
essentially in place when the black hole begins to grow,
and that a dark halo plays negligible role
in this process.  There is ample evidence that the spheroid itself is
triaxial and that the baryonic components are completely dominant
within an effective radius \cite{trim,romeal}.  

I consider the details of gas disk collapse in triaxial
galaxy cores via ``sticky particle'' calculations.
The first problem is that of energy loss;  low angular momentum
gas entering the vicinity of core must dissipate sufficient energy
to fall below the threshold for the disappearance of tube orbits.
I model this process by 
an in-falling gas annulus, with angular velocity insufficient to
balance gravity;  the annulus falls past an equilibrium point and oscillates
radially inward and outward. In multiple
bounces over several dynamical timescales, 
energy is dissipated until a significant
fraction of the gas has penetrated the critical energy
below which tube orbits do not exist.  The gas disc then collapses
to the centre within one or two dynamical timescales.  

Such
a mechanism could not only lead to fuelling of low mass seed black holes 
at a rate near
the Eddington limit, but, as stressed by El-Zant et al. (2003), 
it is also self-limiting.  When the black hole
mass grows to a substantial fraction of the core mass, tube orbits
reappear in the core and the box orbits become chaotic \cite{merrev}.
Since gas is preferentially trapped on non-chaotic orbits, a
more typical gas disc re-emerges with slow, viscosity-driven
accretion onto the SBH.  If core properties are closely
tied to the overall properties of the stellar system, the global
scaling relations for SBHs might be explained.  But it is important
to appreciate that the mechanism described here can only be relevant 
to the early
growth of black holes-- at the epoch of galaxy assembly-- and not
to present activity associated with SBHs in galactic nuclei.

\section{Orbit structure and gas flow in triaxial cores}

Here, following Lake \& Norman 1983 and El-Zant et al. 2003, 
I review the relevance of orbit structure to gas motion
in triaxial systems.
I assume that the spheroidal galaxy, in a principle plane, is
described by the potential
$$\Phi(x,y) = {{{V_o}^2}\over 2}\ln\Bigl({r_c}^2+x^2+{{y^2}\over {q^2}}
 \Bigr)   \eqno(1)$$
where $q\le 1$ (Binney \& Tremaine 1987).  
This logarithmic potential contains a 
simple harmonic core and would, in the case of
axial symmetry ($q=1$), provide a system with a flat rotation curve
(rotation velocity $V_o$) beyond the core.  This corresponds to a
constant density core with a density distribution asymptotically
approaching $1/r^2$ beyond a core radius $r_c$ (similar to an
isothermal sphere).  The central density is given
by
$$\rho(0) = {{3{V_o}^2}\over{4\pi G {r_c}^2}}. \eqno(2)$$

Here, as a numerical example, I take ``typical'' elliptical galaxy
values of $r_c=30$ pc and $V_o = 200$ km/s. In this case,
$\rho(0) = 2.46\times 10^3$ M$_\odot$/pc$^3$ and the total stellar
mass of the core (mass out to $r_c$) would be about $1.5\times 10^8$
M$_\odot$.  In the calculations described below I take $q=0.937$--
a mild deviation from axial symmetry avoiding commensurable oscillations
along the x and y axes.

\begin{figure}
\resizebox{\hsize}{!}{\includegraphics{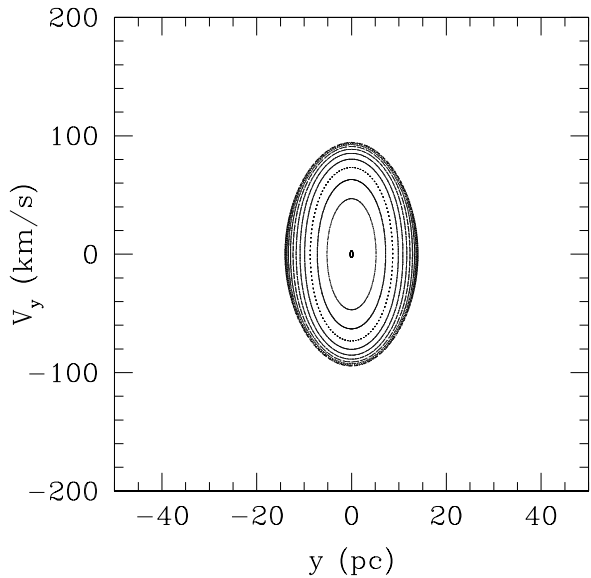}}
\caption[]{Surface of section ($y-\dot{y}$ plane) for an energy
of $1.41\times 10^5$ (km/s)$^2$ in the potential given by eq.\ 1.
This is sufficient to carry particles to a radius of about 
0.5 $r_c$.  The point at the origin is the long axis radial 
orbit and the invariant curves represent boxes librating about this
orbit. This deep in the core, there are no loop orbits.}
\end{figure}

The orbital structure in such a system can be understood by considering
surfaces of section-- the maps generated by progressive penetrations
of a plane in the four 
dimensional phase space by orbits at a given energy. 
Fig.\ 1 is such a surface of section
on the $y$-$\dot{y}$ plane
for orbits at an energy of $1.405 \times 10^5$ (km/s)$^2$ 
-- sufficient to carry the 
particles to a maximum radius of 0.5 $r_c$ (15 pc in this case); i.e.,
these would be orbits fairly deep within the core.  The point at the
centre is the long axis periodic orbit (radial oscillations along the x
axis).  It is evident that all surrounding curves represent
orbits which librate about this long axis radial orbit with no
preferred sense of rotation; there is no integral of motion
analogous to angular momentum, and there are no loop orbits this 
deep in the core.

Fig.\ 2 is a surface of section at a higher energy of $1.45 \times 10^5$
(km/s)$^2$, sufficient to carry a particle to a radius
of 0.75 $r_c$.  By this point the loop orbits, evident as the two sets
of closed curves beyond the box orbits, have developed.  These two
sets correspond to opposite senses of rotation.  The invariant curves
close about two periodic orbits which are slightly elongated
perpendicular to the major axis of the potential distribution.
The loops librate
about these periodic orbits.  Steady state gas flow streamlines
would be expected to correspond to one of these two families
of closed periodic orbits-- the parents of the loop families--
depending upon the sense of rotation.

\begin{figure}
\resizebox{\hsize}{!}{\includegraphics{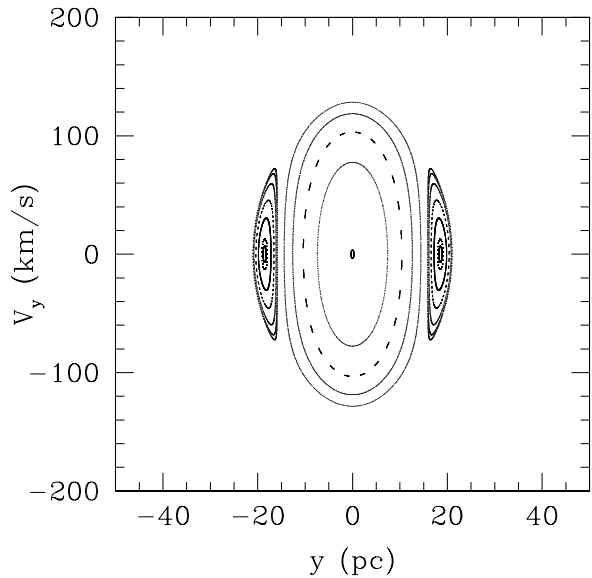}}
\caption[]{Surface of section ($y-\dot{y}$ plane) for an energy
of $1.45\times 10^5$ (km/s)$^2$ in the potential given by eq.\ 1.
This is sufficient to carry particles to a radius of about 
0.75 $r_c$.  Again we see the family of box orbits surrounding
 the long axis radial 
orbit at the origin, but at this energy invariant curves representing
two families of loop orbits, corresponding to different senses of
circulation, have also appeared.}
\end{figure}

At energies lower than $E_T=1.42 \times 10^5$ (km/s)$^2$, 
intermediate
between the two cases shown and sufficient to carry particles out
to a radius of 0.6 $r_c$, there are no loops; i.e., $E_T$ represents
a threshold above which loops are found for this particular
value of $q$ (for smaller $q$ the threshold is at higher energy). 
What, then, is the
fate of gas that diffuses below $E_T$? 
Angular momentum, or at least the integral $I_2$ which
becomes angular momentum in the limit of axial symmetry, is no longer 
conserved;  the only orbits
available are the boxes, and we would expect the gas to collapse to
the centre on a dynamical timescale.  

The expectation is altered, however, by the presence of a
SBH at the centre of the core.  Adding the potential of a point
mass ($-GM_h/r$) to eq.\ 1, where the mass of the black hole, $M_h$,
is a substantial fraction of the core mass, causes the re-appearance
of loop orbits and the re-emergence of the angular momentum-like integral
of motion.  This is illustrated in Fig.\ 5 which is a surface
of section at an energy sufficient to carry particle to a radius
of 0.6$r_c$ when a point mass
of $5\times 10^7$ M$_\odot$ (about 1/3 of the core mass) has been placed 
at the centre of the system described above .  We see now
the presence of two families of loops.
\begin{figure}
\resizebox{\hsize}{!}{\includegraphics{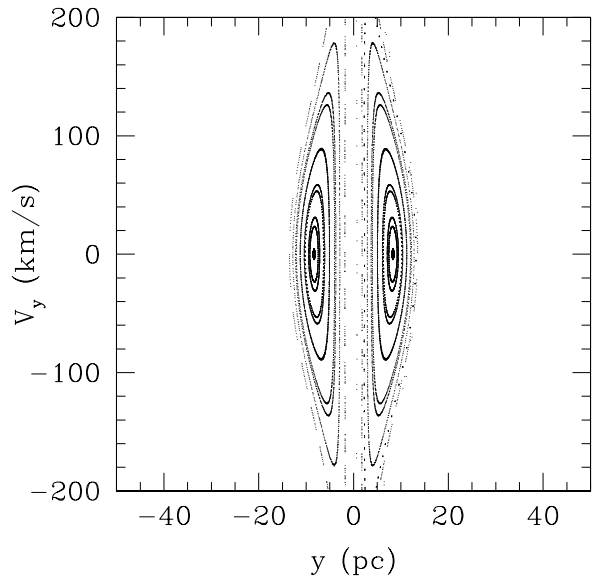}}
\caption[]{Surface of section ($y-\dot{y}$ plane) 
in the potential given by eq.\ 1 to which a point mass of $5\times 10^7$ 
M$_\odot$ has been added.  Here the energy is $1.25\times 10^5$ 
(km/s)$^2$ which, as in the case shown in Fig.\ 1,
is sufficient to carry particles to a radius of about 
0.5 $r_c$.  Here the loop orbits have reappeared deep in the core,
and some fraction of the box orbits have become chaotic.}
\end{figure}
Some fraction of the box orbits become chaotic, but
in most contexts, gas motion in the presence of both chaotic and 
non-intersecting periodic orbits is trapped on the periodic orbits 
\cite{ln83,stva}.  This suggests that the growth of
the black hole by this process of rapid gas accretion in the core would
be self-limiting.  The reappearance of circulating periodic orbits 
forces the flow back to the slow, viscous accretion mode and cuts
off the possibility of further growth at the Eddington limit.
These expectations are supported by 
sticky particle calculations described in the following section.

\section{Sticky particle calculations}

The technique for including dissipation in particle motion 
is essentially that which I have previously applied to simulating 
Galactic Centre clouds (Sanders 1998).  One assumes an interaction
distance $\sigma$ such that initially each particle has about
10 neighbours within this distance.  Then at each time step every
particle adjusts its velocity so as to reduce the velocity difference
with these close neighbours, but only if that velocity difference
is negative (i.e., if the particle are approaching each other).

If $\bf{V_{ij}}$ is the component of the relative velocity along the
line joining the two particles (i.e., the vector at the position of
particle $i$ pointing away from particle $j$), then 
the velocity of particle $i$ during time-step $\Delta t_k$ changes by
$$\delta {{\bf v_{ik}}} = \alpha_k\sum_j^{r_{ij}<\sigma} \bf{V_{ij}}
\eqno(4)$$ 
where
$$\alpha_k = \Delta t_k/\Delta t_d \eqno(5)$$
and $\Delta t_d$ is an adjustable dissipation timescale.  This provides,
in effect, a bulk velocity in which every particle's velocity is adjusted
proportionally to the local velocity divergence, but only if that 
divergence is negative. The algorithm conserves linear momentum and,
in axial symmetry, the angular momentum of an ensemble of particles,
but obviously does not conserve energy.  In the following examples I
take $\sigma = 0.6$ pc and $\Delta t_d = 0.02$ dynamical timescales
(20000 years).

For the mechanism of trapping on box orbits to work, the in-falling
gas disc must not just penetrate to within about 0.5 $r_c$, it
must also dissipate sufficient energy to fall below the threshold
$E_T$ below which no loops are present. Therefore,
as an initial condition I take 4000 ``gas'' particles to be 
uniformly distributed between radii of 20 pc and 40 pc, 
in pure tangential motion about the centre,
but with only 40 \% of the velocity required to balance gravity;
i.e., the centripetal acceleration is 0.16 of the gravitational
acceleration.  
In addition I give the particles a random motion
of about 10 \% of the tangential velocity (10-20 km/s).  This
initial condition is arbitrary, but could correspond to low angular
momentum gas flowing into the core from larger radii-- either as a result
of a stellar merger or as gas lost from stars during normal stellar evolution
(it would seem quite unlikely that gas would arrive in
the vicinity of the core with zero specific angular momentum).
In any case, because the gas comes from further out in the system,
it must dissipate energy before the mechanism I describe can work.

Given this initial condition I consider in-fall 
in three different variants of the potential given by eq.\ 1:
a) The potential is axisymmetric with $q=1$;
b) There is a mild asymmetry with $q= 0.937$ as for the orbits shown
by surface-of-section in Figs. 1-2;  
c) the potential of a point mass of $5\times 10^{7}$
M$_\odot$ has been included in the non-axisymmetric case
as for the surface of section shown in
Fig.\ 3.  The initial distribution of the gas particles and
the final distributions, after eight characteristic orbit times
(8 million years) are shown in Fig.\ 4. 
 
\begin{figure}
\resizebox{\hsize}{!}{\includegraphics{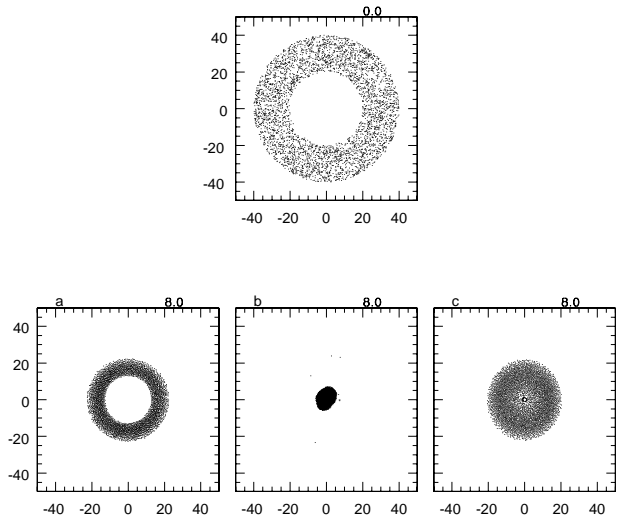}}
\caption[]{The top figure show the initial distribution of 4000
sticky particles for the three cases considered.  The particles are
uniformly distributed between 20 pc and 40 pc but with only 16\%
of the centripetal force required to balance gravity.  The final
distributions, after eight orbital timescales (8 million years) 
are shown below for: a) the axisymmetric potential
($q=0$); b) the non-axisymmetric potential ($q=0.937$); c)
the non-axisymmetric potential to which a central point mass of
$5\times 10^7$ M$_\odot$ has been added. The axes are labelled in pc.}
\end{figure}

Case $a$, inflow in the axisymmetric potential, is interesting 
because it demonstrates how fairly rapid dissipation of energy
can take place.  After eight orbit times we see that 
a ring has formed at mean radius of about 18 pc.  This ring oscillates
initially with large radial excursions
but the oscillations damp away within a few dynamical timescales
due to the effective
dissipation inherent in this sticky particle routine.  
The average orbital energy per particle is shown in Fig.\ 5
(dashed line) where it is evident that the energy decreases
in steps.  These steps down occur when the ring is at its minimum
radius-- at the bounce.  Here the inner part of the ring is 
moving outward while the outer part is still moving inward; hence
there is large compression and dissipation.
The specific angular momentum (average per particle) as a function of
time is shown in Fig.\ 6, and it is clearly conserved to high precision.

\begin{figure}
\resizebox{\hsize}{!}{\includegraphics{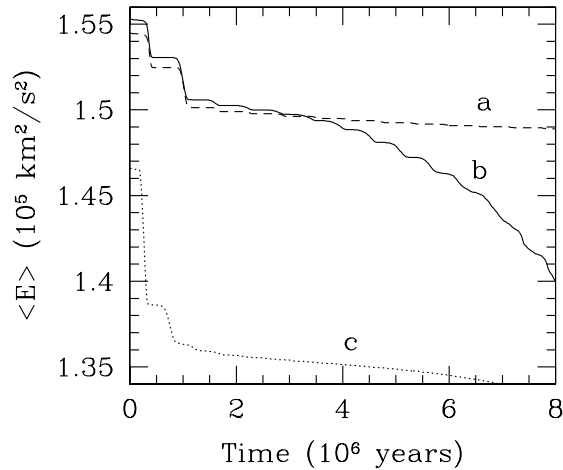}}
\caption[]{The average orbital energy per particle (in $10^5$ km$^2$/s$^2$)
as a function
of time ($10^6$ years) 
for the axisymmetric case $a$ (dashed line) and the non-axisymmetric
case $b$ (solid line), and the non-axisymmetric case with a central
black hole of $5.0\times 10^7$ M$_\odot$, case $c$.
It is evident that the energy decreases in steps
corresponding to maximum compression.  By this process, gas particles
penetrate the energy threshold below which tube orbits disappear and
this leads to the rapid collapse of the gas disc in case $b$.}
\end{figure}

In the mildly non-axisymmetric case $b$,
the simulated gas annulus has collapsed to the centre
within eight rotation periods.  This is because of the
rapid dissipation of energy in radial oscillations-- as is
evident in Fig.\ 5 where again we see the pronounced
steps down in energy corresponding to minimum contraction of
the non-axisymmetric gas ring.  By five or
six rotation periods, most of the gas particles have penetrated
below $E_T$ and entered the inner part of the core 
where tube orbits no longer exist.  The angular momentum,
no longer conserved, then decreases to less than 0.1 of its original value
(Fig.\ 6). Of course, the total 
angular momentum of the entire system of stars and gas must be 
conserved.
This means that the original angular momentum of the gas disk is
is lost to the non-axisymmetric stellar system via torques.  This
would tend to give the trixial system a figure rotation, or, more
likely, would result in a heating of the system and, on the
long term, a restoration of axial symmetry in the core.

\begin{figure}
\resizebox{\hsize}{!}{\includegraphics{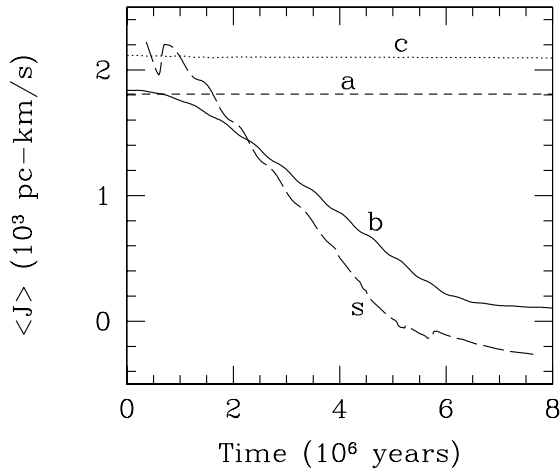}}
\caption[]{The average specific angular momentum (units of $10^3$ pc-km/s)
per particle as a function
of time (orbit times or millions of years) for collapse cases $a$, $b$ and
$c$, as well as for newly formed stars in case $b$ (labelled $s$).  
In the axisymmetric case $a$ the angular momentum is conserved but
in the non-axisymmetric case there is a dramatic decrease as the disc
collapses to the centre.  In the presence of a black hole the angular
momentum is restored as a conserved integral.  With respect to
the newly formed stars, note that this is the average angular momentum per
star; the angular momentum of the initially formed stars is not
lost but is diluted as more stars are formed in the collapse.  The ensemble of
stars formed from the collapsing gas via the compression criterion  
(see eq.\ 5) is characterised, finally, by counter-rotation (negative
average angular momentum).}
\end{figure}

In case $c$, where the black hole has been added, the in-falling
annulus has become a standard accretion disc which very slowly 
drains into the hole.  This is due to the re-emergence of loop
orbits (Fig.\ 3) deep within the core and the presence of a second 
angular momentum-like integral.  Although there is a large dissipation
of orbital energy in the first two bounces (Fig.\ 5),
the angular momentum, after an initial small decrease, is well-conserved
(Fig.\ 6).

The steps by which the gas disc in the non-axisymmetric case $b$ 
collapses is broadly traced by the time sequence shown in Fig.\ 7.
Here it is evident that an asymmetric ring is formed which, after
oscillating in and out, gradually closes and collapses to the centre
\begin{figure}
\resizebox{\hsize}{!}{\includegraphics{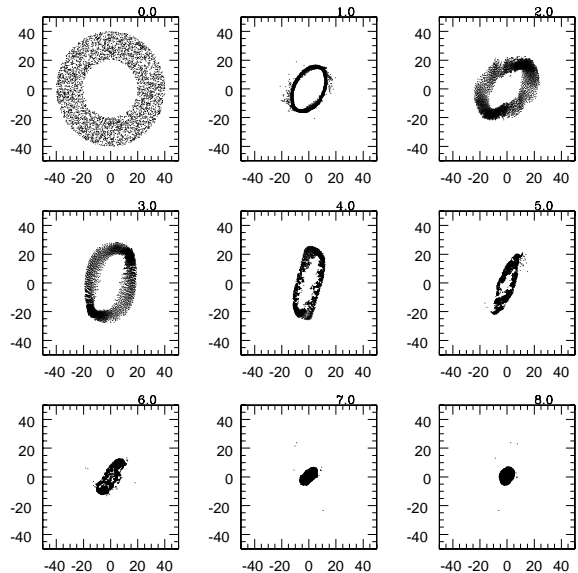}}
\caption[]{A time sequence showing the evolution of the in-falling
gas ring in the mildly non-axisymmetric case $b$.  The frames are
separated by one orbit time or 1 million years (time labelled on upper
right of each frame).  Between frames 5 and 6 the ring completely closes
and the resulting dissipation causes the gas to collapse to the centre.}
\end{figure}
In fact, the time intervals in this figure are too large to show the
details of the collapse; the asymmetric ring opens and closes several
times before the final collapse, but it is clear that the collapse
occurs over several dynamical timescales rather than on
a slow viscous timescale.

During the collapse of the gas annulus, there is significant
compression of the gas and resulting strong shocks.  Therefore, 
star formation would probably proceed in such an environment.
In case $b$, for example, maximum compression occurs at the
point where the elliptical ring has reached its minimum radius, therefore
we might expect star
formation to occur in bursts corresponding to the steps down in 
energy (Fig.\ 5).  Even stronger compression of the gas develops
when the ring collapses to the centre between frames 5 and 6 in
Fig.\ 7 because here a number of the gas particles are on 
a counter-rotating path (recall that the relevant orbits are boxes).

In the sticky particle
technique for including dissipation (eq.\ 4) , the velocity divergence at
time $k$ at the position of particle $i$ can be estimated:
$$ -(\nabla \cdot v)_{ik} = {{|\Delta {\bf v}_{ik}}|\over{\alpha_k\sigma}}
\eqno(5)$$ where $ {\Delta {\bf v}_{ik}}$ is given by eq.\ 3.
To simulate star formation I assume that if the compression defined
by eq.\ 5 exceeds a certain threshold, the dissipation for
that particle is turned off and the particle motion is thereafter
only affected by gravity (as in Sanders 1998).
Here I arbitrarily have set the compression threshold for star formation
at 650 km/(s-pc); with a factor of two 
higher threshold relatively few particles are converted into stars,
and with a factor of two lower threshold, the majority of the 
particles become stars.  With this threshold, about 630 of the
original 4000 gas particle have been converted to stars by the
end of the simulation.

\begin{figure}
\resizebox{\hsize}{!}{\includegraphics{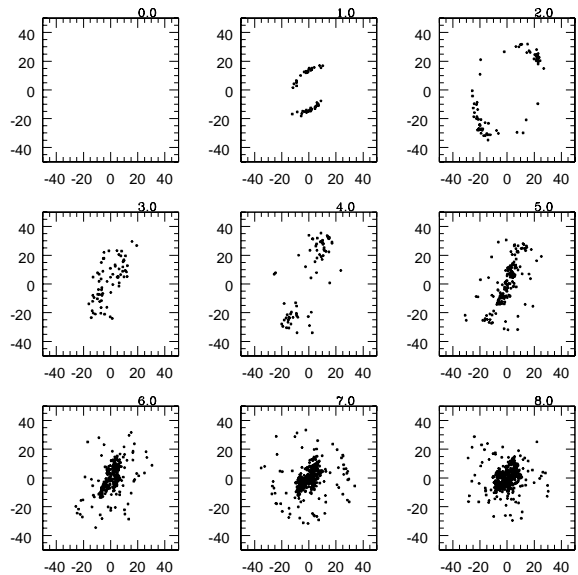}}
\caption[]{A time sequence showing the appearance and distribution
of stars newly formed via compression
in the mildly non-axisymmetric case $b$.  The frames are
separated by one orbit time or 1 million years and correspond exactly
to the frames of the gas distribution shown in Fig. 7.  
Between frames 5 and 6, when the ring completely closes,
the compression is maximum and most stars form.}
\end{figure}

The spatial distribution of these
newly formed stars is shown in Fig.\ 8, which is the same time
sequence as for the gas distribution (Fig.\ 7).  As expected,
the new stars form at maximum compression with a large
burst during the final collapse of the annulus. It is of interest that
the ensemble of stars formed by t=8, primarily in this final
collapse, is counter-rotating with respect
to the original gas disc; this is because those fluid elements which find
themselves moving against the direction of most of the fluid 
experience the largest compression.  The counter rotation is also evident
from Fig.\ 6 where the curve labelled $s$ shows the time dependence of
average angular momentum per star for the newly created stars.  By the
end of the simulation it is negative.  It is also of interest that the
density of these newly formed stars increases toward the centre;
this could create the appearance of a cusp in the presence of a constant
density core. (for movies of all four simulations go to
htpp://www.astro.rug.nl/$\sim$~sanders/movie.html)

\section{A note on scaling relations}

Any model for the growth of black holes in galactic nuclei must address
the issue of the surprisingly tight relation between the black hole
mass and the larger scale velocity dispersion in the spheroidal
galaxy; i.e.,
$$ M_h \propto {V_s}^\beta \eqno(6)$$
where $\beta$ is in the range of 4 to 4.5 \cite{fm,nuk1,trem}.
This is possible in the context of the mechanism described here 
because it is self-limiting with the rapid growth of the SBH occurring
until the hole mass becomes on the order of 10\% to 20\% of the core mass.
A correlation between the core mass and the velocity dispersion
is the remaining missing ingredient. 

This may be possible if the formation of a core is viewed in terms
of a maximum phase space density \cite{dalhog}.  The average
density in a core giving rise to the potential described
by eq.\ 1 is $$\bar{\rho_c} \approx {{2{V_s}^2}\over{2\pi G{r_c}^2}}
\eqno(7)$$ which means that the average phase space density 
$\bar{\rho_c}/{V_s}^3$ is 
$$f_c= {3\over {2\pi G {V_s} {r_c}^2}}\eqno(8)$$
Combining these relations (given that the core mass is $\rho_c{r_c}^3$)
we find
$$M_c\approx {\Bigl({{V_s}\over G}\Bigr)}^{3\over 2} {{f_c}^{-{1\over 2}}};
\eqno(9)$$ i.e., there is a built-in correlation between core mass
and velocity dispersion.

It has been demonstrated that the average course-grained
phase space density in galaxies decreases as galaxy luminosity increases
\cite{herneal,dalhog}.
If spheroidal galaxies are more or less homologous then,
combined with a luminosity-velocity dispersion relation (Faber-Jackson),
this would imply $f_c \approx {V_s}^{-p}$ where $p$ is between
4 and 6.  Then combined with eq.\ 9 we have
$$M_h \approx 0.1 M_c \propto {V_s}^{(3+p)/2}$$ which could explain
the observed relation eq.\ 6.  This is all quite speculative and 
it is unclear that such correlations could account for the tightness
of the $M_h-V_s$ relation.  

It also entirely depends upon the pre-existence of cores in
spheroidal galaxies-- cores with properties set by phase space
constraints.  In pure CDM halos, of course, there are no
cores because there are no phase space constraints, but there are
clear observational indications that cores do exist at least in dwarf spirals
\cite{dbeal,salref}.  This issue remains to be settled.

\section{Summary and speculation}

Observed constant density (or constant surface brightness) cores in
early type galaxies range in radius from 10 to 1000 pc \cite{fabereal}.  
Gas deposited in such
cores from some exterior source, must collapse into the centre
if the core is mildly non-axisymmetric with a non-rotating
figure and not yet dominated by the central black hole.  
This collapse is due to the absence of simple closed, 
non-self-intersecting periodic orbits below a threshold energy;
such orbits normally play the role of 
attractors in the phase space of the
dissipational medium.  In some inner fraction of the core, 
which can be significant even in the presence of a weak deviation 
from axial symmetry,
there are only box orbits librating about the long and short axis radial
orbits.  

Once the gas, through dissipation, has penetrated below the
threshold energy,  collapse is rapid-- occurring within a few
dynamical timescales, and this could lead to Eddington growth of
small seed black holes.  Because, in such dynamical collapse, 
strong shocks develop in the gas  with resulting high compression,
some fraction of a collapsing gas disc 
may be turned into stars; the density of these new stars increases into
the centre and may disguise the original constant density core.
In this sense, it is interesting that bright nuclei are found in
29\% of cored galxies and 60\% of galaxies with central powerlaw cusps 
\cite{lauer5}, and that these cusp nuclei are significantly bluer than the
surrounding galaxy.

The process is self-limiting in the sense that when the mass
of a central black hole has grown to be 10\% to 20\% of the 
mass of the stellar core, the simple periodic orbits reappear within the
core; angular momentum re-emerges as an integral of motion. 
Matter will continue to flow
into the black hole but on a much longer viscous time scale.
Such a mechanism would curtail the rapid growth of the black
hole beyond a fraction of the core mass and thus limit the present
observed mass of nuclear black holes.  If there is a correlation between
the initial properties of the core and the global properties of
the galaxy, such as a correlation between core radius and effective
radius of the spheroid \cite{fabereal}, then the global scaling relations
for SBHs-- e.g., the mass-velocity dispersion relation-- 
might be understood in the context of this mechanism.

The crucial question is whether or not constant density cores exist
initially in spheroidal stellar systems.  There are various observations
consistent with the idea
that the central density distributions are initially cusp-like and it is
the orbital decay of massive binary black holes that creates a
core-- e.g., core galaxies are rounder and 
have reduced colour gradients \cite{lauer5}.  On the other hand,
we have seen that star formation expected in the rapid collapse 
of a central gas disc could produce both an apparent
cusp in the density distribution and a colour gradient.
It would appear that the issue of initial cores is not yet settled,
but the mechanism discussed here could contribute to rapid fuelling of a
black hole in any constant density triaxial core so long as it is
not gravitationally dominated by the black hole.  The simulations
would, for example, also be relevant to the model of El-Zant et al.
(2003) where cores are formed in a triaxial CDM halo during the
initial baryonic collapse.

There are several possible sources for in-falling gas.  
One favoured possibility is that low angular momentum
gas is supplied to the central region in merger events.  The
presence of counter-rotating discs in the central regions
of some galaxies would seem to support this scenario.
However, we have seen that a counter-rotating stellar disk can also be 
formed
by the mechanism described here, even though the gas originally
shares the rotation of the galaxy at large.  This would 
give weight to a second possibility:  low angular momentum 
gas can be supplied via mass loss from the stars in a slowly
rotating system.  Some of this gas is blown away in supernova
heated winds, but some fraction of the gas may cool and flow into
the centre, either simultaneously with a hot wind or in an
unsteady cooling flow alternating with hot wind phases \cite{co}.  
In either case-- for an external or internal source--, the
mechanism discussed here would bridge the final gap for transfer of 
gas from the core boundary to the black hole. It is a significant
gap; gas in equilibrium at the core boundary must reduce its angular 
momentum by a factor of $10^{-8}$ to arrive at the Schwarzschild radius.  

It has recently been discovered
that the young stars observed within the central parsec of the Galaxy
lie in two distinct disc systems at large inclinations to each other
and with different senses of rotation \cite{paum}.  The mechanism
of dynamical collapse of gas discs would not apply here because
the central parsec is gravitationally dominated by the 
$3\times 10^6$ M$_\odot$ black hole; however, an in-falling gas disc
in the presence of a dominant black hole
(as in case $c$ above), also bounces several
times before settling into an accretion disk.  It is evident in
Fig.\ 5 that there is a large dissipation of energy in the
first two bounces-- a dissipation corresponding to strong 
compression. Thus it is possible that in the bounce
star formation could proceed through strong shocks even in
the near presence of the central black hole.  This is a topic for
future consideration.
 
I am grateful to Colin Norman for rekindling my interest in this problem.
I also thank Scott Trager for useful conversations about merging and
for assistance in producing movies of these
simulations.  Finally, I am very grateful to Isaac Shlosman for
bringing the work of El-Zant et al. to my attention,
and I apologize to the authors for not citing this most relevant
paper in the previous version.


\begin{thebibliography}{}
\bibitem [Bahcall \& Wolf 1976] {bw} Bahcall J.N., Wolf R.A., 1976,
  ApJ, 209, 214
\bibitem [Binney \& Tremaine 1987] {bintre} Binney J., Tremaine S
  1987, {\it Galactic Dynamics}, Princeton Univ. Press, Princeton
  N.J. 
\bibitem [Ciotti \& Ostriker 1997] {co} Ciotti L., Ostriker J.P.,
  1997, ApJ, 497, L105
\bibitem [Dalcanton \& Hogan 2001] {dalhog} Dalcanton JE,
  Hogan CJ, 2001, ApJ, 561, 35
\bibitem [de Blok et al. 2001] {dbeal} de Blok WJG, McGaugh SS,
  Bosma A, Rubin VC, 2001, ApJ, 552, L23
\bibitem [Elvis, Risaliti \& Zamorani 2002]{elvis} Elvis M., 
  Risaliti G., Zamorani G., 2002, ApJ, 565, L75 
\bibitem [El-Zant, Shlosman \& Hoffman 2001] {ezsh} El-Zant AA, Shlosman
  I, Hoffman Y, 2001, ApJ, 560, 636
\bibitem [El-Zant et al. 2003]{ezeal} El-Zant AA, Shlosman I,
  Begelman MC, Frank J, 2003, ApJ, 590, 641
\bibitem [Eyles et al. 2006] {hred2} Eyles LP, et al., 2006, 
  astro-ph/0607306 
\bibitem [Faber et al. 1997] {fabereal} Faber S.M. et al. 1997,
  AJ, 114, 1771
\bibitem [Fabian \& Iwasawa 1999] {fabiw} Fabian A.C., Iwasawa K., 1999,
  MNRAS, 303, L34
\bibitem [Ferrarase \& Merritt 2000] {fm} Ferrarase L., Merritt D., 2000,
  ApJ, 539, L9
\bibitem [Gebhardt et al. 2000] {nuk1} Gebhardt et al., 2000,
  AJ, 119, 1157
\bibitem [Gentile et al. 2004] {salref} Gentile G, Salucci P, Klein U, 
  Vergani D, Kalberla P, 2004, MNRAS, 351, 903 
\bibitem [Hernquist, Spergel \& Heyl 1993] {herneal} Hernquist L,
  Spergel DN, Heyl JS, 1993, ApJ, 416, 415
\bibitem [Iye et al. 2006]{hred1} Iye M, et al., 2006, Nature,
  443,186
\bibitem [Lake \& Norman 1983]{ln83} Lake G., Norman C., 1983, ApJ,
  270, 51
\bibitem [Lauer et al. 2005] {lauer5} Lauer T.R. et al., 2005,
  AJ, 129, 2138
\bibitem [Merritt 2006] {merrev} Merritt D., 2006, Rep.Prog.Phys. D69
  2513
\bibitem [Mirosavljvi\'c \& Merritt 2001] {mm} Mirosavljvi\'c M.
  \& Merritt, D., 2001, ApJ, 563, 34
\bibitem [Peebles 1972]{peeb} Peebles PJE, 1972, Gen.Rel.Grav., 3, 61
\bibitem [Paumard et al. 2006] {paum} Paumard T. et al., 2006,
  ApJ, 643, 1011
\bibitem [Romanowsky et al. 2003] {romeal} Romanowsky A, et al. 2003,
   Science, 301, 1696
\bibitem [Sanders 1998] {rhs98} Sanders RH, 1998, MNRAS, 294, 35
\bibitem [Sanders, Teuben \& van Albada 1983] {stva} Sanders RH,
  Teuben PJ, van Albada GD, 1983, in {it\ Interenal Kinematics and
  Dynamics of Galaxies}, IAU Symp.100, ed. L. Athanassoula,
  (Reidel, Dordrecht), p.221
\bibitem [Schwarzschild 1979] {schw} Schwarzschild M, 1979, ApJ, 232, 
  236 
\bibitem [Sellwood \& Moore 1999] {sellmoore} Sellwood JA, Moore EM,
  ApJ, 510, 125
\bibitem [Shaw et al. 1995] {shaw} Shaw M, Axon D, Probst R,
  Gatley I, 1995, MNRAS, 274, 369
\bibitem [Shlosman, Begelman \& Frank 1990] {shlos} Shlosman I., Begelman M.C.,
  Frank J., 1990, Nature, 345, 679 
\bibitem [Tremaine et al. 2002] {trem} Tremaine S et al., 2002, ApJ, 574, 740
\bibitem [Trimblay \& Merritt 1995] {trim} Trimblay B, Merritt, D, 1996,
  AJ, 110, 1039 
\bibitem [Wall et al. 2005] {weal} Wall J.V., Jackson C.A., Shaver P.A.,
  Hook I.M., Kellerman K.I., 2005, AA, 434, 133
\bibitem [Young 1980]{yng} Young P., 1980, ApJ, 242, 1232

\end{thebibliography}
\end{document}